\documentclass[sigconf, preprint]{acmart}

\def\BibTeX{{\rm B\kern-.05em{\sc i\kern-.025em b}\kern-.08emT\kern-.1667em\lower.7ex\hbox{E}\kern-.125emX}}

\usepackage{xcolor}
\usepackage{graphicx}
\usepackage{hyperref}
\usepackage{wrapfig}
\usepackage{multirow}
\usepackage{listings}
\usepackage{tabularx}
\usepackage{geometry}
\usepackage{longtable}
\usepackage{enumitem}

\definecolor{lgray}{rgb}{0.9, 0.9, 0.9}

\setcopyright{none}
\renewcommand\footnotetextcopyrightpermission[1]{} 
\usepackage{fancyhdr}
\begin{document}

\title{(Position Paper) Privacy with Surgical Robotics: Challenges in applying contextual privacy theory}

\author{Ryan Shah}
\email{ryan.shah@strath.ac.uk}
\affiliation{
 \institution{University of Strathclyde}
}

\author{Shishir Nagaraja}
\email{shishir.nagaraja@strath.ac.uk}
\affiliation{
 {\normalsize (Corresponding Author)}\\
 \institution{University of Strathclyde}
}

\begin{abstract}
  The use of connected surgical robotics to automate medical
  procedures presents new privacy challenges. We argue that
  conventional patient consent protocols will no longer work. Indeed
  robots that replace surgeons take on an extraordinary level of
  responsibility. Human surgeons undergo years of training and peer
  review in an strongly regulated environment, and they derive trust
  via a patient's faith in the hospital system beyond the surgeon
  him/herself. Robots on the other hand derive trust not through the
  hospital system they operate within, but the integrity of the
  software that governs their operation. From a privacy perspective,
  there are two fundamental shifts. First, the threat model has
  shifted from one where the humans involved were untrusted to one
  where the robotic software is untrusted. Second, the basic unit of
  privacy control is no longer a medical record, but is replaced by
  four new basic units: the subject on which the robot is taking
  action; the tools used by the robot and the valid form of tool
  usage; the sensors (i.e data) the robot can access and the
  operations permitted over them; and, finally access to monitoring
  and calibration services which afford correct operation of the
  robot. We suggest that contextual privacy provides useful
  theoretical tools to solve the privacy problems posed by surgical
  robots. However, it also poses some challenges: not least that the
  complexity of the contextual-privacy policies, if rigorously
  specified to achieve verification and enforceability, will be
  exceedingly high to directly expose to human
  regulators/auditors/users that review contextual privacy
  policies. Another key challenge, is that contextual privacy
  accommodates information flows but not property. A medical robot
  works with both information and physical tissues and fluid
  material. While informational norms allow for judgements about
  contextual integrity and the transmission principle governs the
  constraints applied on information transfer, nothing is said about
  material property. Certainly, contextual privacy provides an anchor
  for useful notions of privacy in this scenario. However, it appears
  the to solve the privacy challenges in the field of surgical
  robotics well, one needs to consider extending contextual privacy to
  cover both information and material flows.
\end{abstract}

\maketitle

\section{Introduction}
\label{sec:introduction}

Modern robotics has transformed a range of application areas, including
but not limited to surgical robotics, industrial robotics and autonomous
vehicles. The implementation of robotics in these areas brings forth the
potential to increase the efficiency of output, as well as the accuracy
of operations amongst many other things. In the case for surgical robotics,
a high level of accuracy and precision is key, which in a surgical
environment could mean the difference between life and death. Connected
surgical robots, such as the da Vinci surgical system
~\cite{talamini2003prospective}, are increasingly playing a leading role
in carrying out teleoperated surgical procedures, including: performing
incisions, controlling blood loss, and bone milling, under the oversight
of human surgeons. Likewise, the majority of teleoperated industrial robot
architectures are comparatively similar to teleoperated surgical robots
~\cite{quarta2017experimental}. Industrial robots accompany a large
portion of installed robotics systems, which range from aiding in collection, packaging, and distribution of produce, to the automotive industry.

Surgical robotics as a technology piece will have a transformative effect on medical health and well being. However, its security and privacy implications are yet to even be roughly ascertained. As we start using connected robotics in medical contexts, there will be a few challenges.

Conventionally, patient privacy in the context of medical procedures is based on the consent protocols that have evolved over the centuries. These protocols are based on patient's faith in the system, to train surgeons to a high standard starting. Surgeons start as apprentices and make their way through a systematic process that retains patient trust, by placing patient well being at the centre of its focus. The question from a privacy consent perspective is, can surgical robots deliver similarly well?

\section{Privacy challenges in surgical robotics}

To answer the question at the end of the previous section, we can
start by examining how existing mechanisms of privacy control would
work if robots replace human surgeons. Our first observation is that a
robot does not simply replace a human surgeon, but brings a new
ecosystem of players into the ring. For instance, in order to operate
correctly, calibration data flows from a National Measurement
Institute (such as NIST) via intermediate calibration agencies, to the
surgical robot located in a hospital. Such a step would be necessary
to ensure that the robots sensors are properly calibrated in advance
of a medical procedure (information flow from a high-integrity source
to low-integrity destination). The traffic around calibration
activities is procedure dependent, and likely to expose private
information about the patient up the chain of service providers who
collectively provide an on-the-fly calibration service. This is often
required in the event of invalid calibration being detected during the
surgical procedure, resulting in the accuracy and precision being
questioned, and to this we must perform on-the-fly calibration.
Beyond calibration, analytics and system monitoring data flows from
the robot to the manufacturer, and possibly the hospital's insurer and
the medical regulator who have a stake in it operating correctly.

The intuitive notions of multi-level and compartmented access clearly
apply to this problem. Unfortunately, the confidentiality-and-privacy
preserving mechanisms used to implement these notions were developed
when the granularity of control was at the level of a single
file. However, the granularity of control required in a surgical robot
for the purposes of calibration is not at the level of a file read
from a remote file system. What then is the basic unit at which
patient consent should be exercised?

Instead of considering patient consent over files/records, we need to
start thinking about patient consent over these newer basic units of
control. If data, services, operations, and most importantly the
subject themselves are not appropriately managed, then privacy
compromises will follow. We propose four types of basic units over
which privacy controls must be applied:

\begin{enumerate}
	
	\item {\bf Data:} The sensors (and corresponding data) the robot can access, and the operations permitted over them
	\begin{itemize}
		\item During a surgical procedure, some sensors would not be required. For example, a temperature sensor would likely not be required unless suturing was necessary in a procedure. Thus, access to this sensor should not be permitted. Similarly, the data output from sensors should be controlled, as it could leak information about the patient's condition.
	\end{itemize}		
	
      \item {\bf Services:} Access to image-analysis-as-a-service,
        machine-learning-as-a-service, among others. Additionally,
        monitoring and calibration services which enable accurate and
        precise robot operations.
	\begin{itemize}
		\item We should restrict access to the monitoring
                  services over what they should be monitoring. For
                  example, monitoring who calibrates what robot should
                  only be performed at the same level as the
                  calibration taking place, and reading of logged
                  monitoring data should be restricted to the same
                  level as calibration. Keeping these restrictions in
                  line across the calibration information flow would
                  aid in avoiding leakage across levels. For example,
                  information about the specific procedure may leak
                  via: the reference data used for calibration, meta
                  information about the calibrated devices, and the
                  conflicts of interests raised by the aggregation of
                  meta calibration-information (at a calibration
                  service provider who is servicing end-users in a
                  situation of conflict).
	\end{itemize}

	\item {\bf Subject:} on which the robot is performing a procedure
	\begin{itemize}
	\item In the case for surgical robotics, the subject would be
          the patient undergoing a surgical procedure which is carried
          out solely by the robot, by a robot under human operation or
          alongside a human (cobot). The subject is not data,
          information, a file, or a service; it's the whole or part of
          a human or animal body. As an example, waste material from a
          procedure can lead to privacy violations if they are
          labelled as non-human waste and subsequently be analysed to
          violate patient privacy. 
 	\end{itemize}
	
	\item {\bf Operations:} The valid forms of tool usage in the
          context of the procedure (surgical robots can often access
          an array of tools using six to eight arms).
	\begin{itemize}
		\item If a surgical robot is performing a colonoscopy,
                  it should only allow equipment to be used relative
                  to the procedure. For example, a colonoscope
                  (flexible tube with CCD camera attached) is an
                  allowed tool, used to view the patient's colon. A
                  scalpel in this case should not be allowed, as it is
                  unlikely that any incisions would be made in this
                  procedure. Similarly, we must ensure that the
                  colonoscope is used properly, to minimise risk of
                  abdominal distension and rare cases of
                  tears/bleeding.
	\end{itemize}

\end{enumerate}

We argue that patient consent protocols will see a significant change
due to the introduction of surgical robotics. In comparison with human
surgeons, there are two key differences: first, the untrusted party is
the software running on the robot rather than the human; and second,
the conventional target of privacy protection is replaced by four
distinct targets none of which is a file or database record. With a
change as fundamental as that it follows that the conventional consent
protocols and mechanisms face a redesign. The next section explains
how this might be done as an application of contextual privacy.

\section{Contextual privacy theory and surgical robots}
Clearly, in the context of surgical robotics, notice and consent will
not work. The dominant approach of delegation to a trustworthy party
(human surgeon) is also a moot suggestion. Therefore, can contextual privacy
theory provide the tools to solve this problem? One suggestion within
the community has been to replace consent protocols with context-based
privacy policies. We argue that this would better reflect the values
of keeping patient's privacy and dignity, as well as their well being
at the centre of the connected medical environment, rather than an
economic view of data production.

Contextual integrity~\cite{nissenbaum2004privacy} is a theory of
privacy that links the protection of patient data to the norms of
personal information flow within specific contexts. This theory argues
that ``informational norms specific to particular contexts govern the
flow of personal information from one entity to another''. Indeed the
theory of contextual integrity provides a contextual framework for
evaluating the flow of personal information between agents within the
ecosystem of surgical robotics to help identify and verify whether
certain patterns of information flow are acceptable within the context
of the surgical procedure.

Within her conceptual framework, Nissenbaum~\cite{nissenbaum2004privacy} proposed several principles of privacy. The first principle says that privacy is provided by appropriate information flows.  Let us examine whether this is helpful in our context. In the calibration hierarchy, information flows from the top level NMIs to the robots at hospitals, situated at the lowest level. As described earlier, the integrity is highest at the top level and lowest the bottom level. Thus, we should forbid the hospital level to write information to the top level. Likewise, we define that the writing of information can only flow downwards. In terms of patient privacy, we should only allow the reading of information to flow upwards. If higher levels were granted access to read information from the hospital level, this would leak information about the robot's calibration. In the case for on-the-fly calibration, reading the information output from the calibration process from on-the-fly calibration for a specific surgical procedure would lead to leakage of information surrounding the surgical procedure. Similarly, combining this with patient admission and discharge times, and other information would compromise the privacy of patients.

The second principle of contextual privacy is the development of appropriate norms for enforcing privacy controls. By providing norms, such as ``a calibration engineer can only calibrate robots at a single hospital, under conditions of strict confidentiality, to avoid leaking information to other hospitals'', we can evaluate hypothetical scenarios to examine whether it violates contextual privacy. We can supply many hypothetical cases to be evaluated against a single norm. For example ``Engineer X can calibrate robot A and B at hospitals 1 and 2, respectively'', is clearly a violation of contextual integrity against the norm that a single engineer can only calibrate robots at a single hospital.

%

\subsection{Challenges in applying contextual privacy theory}
The challenge of managing the complexity of these contextual-privacy
policies and analysing them should not be underestimated. It seems
clear that expressing privacy policies in a way that rigorously
conveys the risks and benefits to patients so as to be enforceable, in
a verifiable manner, will be too technical to expose patients
to. Particularly as they are unlikely to fully grasp the details given
the context of their impending medical procedure. The average patient
will likely fair poorly when assessing the impact of infection risks,
probabilities of life threatening complications, and compromised or
failed medical procedures arising out of inadequate access control of
a patient's internal organ. Most of us are simply too ignorant about
our physical anatomy to make proper assessments and instead rightly
place our faith in the system of medical practice. Indeed, prospective
surgeons undertake long years of study as interns and apprentices,
undergo frequent certifications, peer review, experience regulation,
and comply with a professional code. Above all, patients trust is
based on an expectation of a service that promotes human welfare and
well being from the surgical staff and large medical facilities. This
faith in the system underpins access control protocols that, albeit
imperfect, have socially evolved over several hundred years. In
contrast, access control protocols on robots must carry the collective
weight of all these expectations.

Zimmer~\cite{zimmer2008privacy} describes norms of behaviour to vary dependent on particular contexts. The norms can govern the flow of patient information in particular contexts, such as what type of information as well as the quantity of information that is relevant to a particular surgical procedure. Informational norms consist of three agents that are relevant to our case: the information subject, storage and distributor of information, and the information receiver. In the case for surgical robots which interact with vulnerable patients, the agents vary upon context. For example, in a surgery involving a teleoperated robot, one surgical command (i.e. rotating the actuator by 5 degrees) can involve more than one instance of an agent. A surgeon would distribute the command to the robot, who in turn may record this command in a cache to be later stored, or may send it to a logging system to be recorded. Similarly, the information subjects would include the robot and surgeon, as well as the patient being operated on. Thus, we also question whether defining a policy for a single robot movement would become too complex for a patient to understand. We can argue that we can define policies to cover an entire surgical procedure, however with many informational norms being evaluated that may consist of more than one instance of an agent, the patient may find it far too complex to interpret.

Given that that expressing consent-based privacy policies in a way
that is rigorous enough to make them verifiable and enforceable is
going to be too technical to expose end users to directly. This
suggests that there is going to be a role for expert review of
policies. It also suggests that delegation to experts needs to be done
to a standard that retains the faith that patients have conventionally
placed in human surgeons. A complimentary approach is the development
of appropriate formal analysis to mitigate the complexity. While
verification techniques based on formal methods might help, we believe
a complete solution must involve experts to provide some level of
governance over these new forms of privacy controls in connected
medical environments.

The surgical robot is a human facing cyber-physical system, so it
doesn't just work with information, it also works with physical
objects -- body parts and human tissue. Privacy constraints must
therefore be applied not just over information flows but also the
physical material the robot is handling. From a consent perspective,
in English and Scottish Law, tissue removed during a surgical
procedure with due consent is considered ``abandoned'' by the patient
and can legally be used for research purposes (Human Tissue Act
2004). However, activities such as clinical audit, performance
assessments, and quality assurance require further explicit consent. A
partially compromised surgical robot could potentially mislabel human
tissue collected after a procedure as research tissue.  Instead of
being incenerated, when such tissue is stored and subjected to
unauthorised commercial exploitation or analysis then privacy
violations will potentially occur (in abundance). The main point we
are making here is that privacy considerations apply not just to
information (data and services), but to physical ``property'' as
well. And, therein lies a basic challenge with applying any
information-centric privacy framework, including contextual privacy,
to the problem domain of surgical robotics. 

\subsection{Applying Contextual Privacy to Policies}

As previously described, managing the complexity of applying contextual privacy to policies for evaluation, whilst ensuring that they can be expressed rigorously in a way to not be exposed too technically to patients, is a challenge. The categorisation of informational norms may vary, from virtually variation to being completely different, dependent on the context. For example, defined informational norms for a minimally invasive procedure would vary a fair amount in comparison to a fully invasive procedure.

To understand the complexity of applying contextual privacy for policies, ideally, we should evaluate patient experiences with understanding surgical procedures from talking with their doctors/surgeons. These can then be formulated into the appropriate norms for specific surgical contexts, and then be applied to a standard which retains conventional patient-surgeon faith. However, with the option currently unavailable, we can define realistic example constructs to examine such a complexity, as shown in Table~\ref{table:norms}.


\bgroup

\scriptsize
\def\arraystretch{2.5}
\setlength\tabcolsep{12pt}
\setlength\LTleft{0pt}
\setlength\LTright{0pt}

\begin{table*}[t]
	\centering
	\begin{tabularx}{\linewidth}{ |X|X|X|X| }
  		\hline
    	Context & Sender \& Receiver & Transmission Principle & Example Norm(s) \\
    	\hline
    	Initial Calibration & Manufacturer -- receives 
		meta-information about the surgical procedure via monitoring information about 
		wear-and-tear status of all stepper motors, arms, and sensors.\par
		Calibration agencies and NMIs -- receive meta-information about the 
		procedure via relevant sensors used in the procedure. & Information divulged to manufacturer and 
		calibration agencies organically by the process of maintaining high 
		physical integrity and the accuracy of sensors. & Robot components must have valid and verified calibration, to ensure a high degree of accuracy and precision. \\
		\hline
		
		On-the-Fly Calibration & Calibration Facilities -- notify subsequent levels up to the robot's components of invalid calibration at a particular level in the calibration hierarchy.\par Surgeon -- informed of invalid calibration during surgery and should make a call for on-the-fly calibration, dependent on the context, to minimise risk to the patient. & Information flows from the top level of the calibration hierarchy (NMIs) towards the hospital in a series of first-hop events. If invalid calibration was detected at an NMI, they would inform the first subsequent units at the next level, and so forth until the surgical robots at the hospital level. & During a surgical procedure, surgical robots must retain valid calibration to maintain a high degree of accuracy and precision. If the calibration is deemed invalid, the robot must be recalibrated on-the-fly. \\
		\hline
		
		Community (Group) Calibration & Calibrator -- wants to calibrate a robot component.\par Robot -- neighbouring components in the robot system will form a group decision to permit/deny calibration. & Neighbouring components will use a form of signalling, possibly through a bus link between them (CAN), to make decisions with other components. & To avoid constant recalibration requests, which can leak sensitive information about the robot operations and its calibration (and possibly the patient), other components must learn to make consensus decisions to authorise recalibration. \\
		\hline
		
		Teleoperated Surgical Procedures & Surgeon -- sends commands/movements via a surgeon's console to the robot.\par Robot -- receives commands from the surgeon's console. & Commands will be transmitted via hardware cable between the surgeon's console and the robot or in some cases, wirelessly. & The robot will only accept incoming commands from the surgeon's console, in either transmission case, and no other sender. \\
		\hline
		
		Autonomous Surgical Procedures & Robot -- receives data from a reliably trained and trustworthy learning mechanism, to make decisions on movements and operations to perform.\par Robot -- sends movements and commands to embedded controller to move the robot. & The robot will receive data and send commands through its internal network, such as via a controller area network (CAN)~\cite{farsi1999overview}. &
		\begin{enumerate}[leftmargin=*]
			\item The robot should only accept data input from a verified classifier that is not prone to leaking information~\cite{hitaj2017deep,hospodar2011machine}.
			\item The learning mechanism to provide data input should be trained and tested from reliable sources (i.e. a human surgeon).
		\end{enumerate} \\
		\hline

	\end{tabularx}
	
	\vspace{0.5cm}
	\caption{Table of Norms for Estimating Complexity}
	\label{table:norms}
\end{table*}

\egroup

\normalsize
It is clear that we can define norms to express contextual privacy in a way that can be easily understood by patients. Trust can be inferred by having a surgeon's input for training autonomous surgical robots, as well defining norms which can envelop patient's trust by ensuring reliable calibration leading to maintained robot accuracy. However, for each of the 5 contexts we present norms for, there can be many different norms that can cover all aspects relevant to the context. As well as this, from a point of evaluation, discussing contexts and norms with real surgeons and calibration agencies would provide more in-depth details that we can not simply provide examples for. We show that the complexity can be minimised using the norms as shown in Table~\ref{table:norms}, but a more in-depth evaluation of these with input from hospitals and calibration agencies will allow us to further examine how we can manage complexity and privacy for real case examples.

\section{Conclusion}
Contextual privacy theory provided a useful conceptual framework within which to situate privacy policies for surgical robots. It also presented three challenges. The main  challenge was of expressing policies that were rigorous enough to be verifiable, and yet simple enough to be understood by patients. A further basic challenge was that information flow constraints have traditionally been expressed where the level of control is a file, however with surgical robots we have four different basic units of control: the subject, operational tool set, data, and services. The final challenge we faced was how to deal with physical material (human tissue), using a consent framework designed to constrain information flows. We don't expect any easy solutions to these challenges.

\bibliographystyle{ACM-Reference-Format}
\bibliography{references}




\end{document}